\documentclass{pnastwo2}

\usepackage{graphicx}
\usepackage{pnastwof2}
\usepackage{amssymb,amsfonts,amsmath}
\usepackage{cite}
\usepackage{balance}

\begin{document}

\title{Neural Dust: An Ultrasonic, Low Power Solution for Chronic Brain-Machine Interfaces}

\author{Dongjin Seo\affil{1}{Department of Electrical Engineering and Computer Sciences and }, Jose M. Carmena\affil{1}{}\affil{2}{Helen Wills Neuroscience Institute, University of California, Berkeley, CA, USA}, Jan M. Rabaey\affil{1}{}, Elad Alon\affil{1}{}\affil{3}{Joint senior authors},
\and
Michel M. Maharbiz\affil{1}{}\affil{3}{}}

\contributor{Correspondence to: djseo@eecs.berkeley.edu}

\maketitle

\begin{article}
\begin{abstract}
A major hurdle in brain-machine interfaces (BMI) is the lack of an implantable neural interface system that remains viable for a lifetime. This paper explores the fundamental system design trade-offs and ultimate size, power, and bandwidth scaling limits of neural recording systems built from low-power CMOS circuitry coupled with ultrasonic power delivery and backscatter communication. In particular, we propose an ultra-miniature as well as extremely compliant system that enables massive scaling in the number of neural recordings from the brain while providing a path towards truly chronic BMI. These goals are achieved via two fundamental technology innovations: 1) thousands of 10 -- 100 $\mu$m scale, free-floating, independent sensor nodes, or neural dust, that detect and report local extracellular electrophysiological data, and 2) a sub-cranial interrogator that establishes power and communication links with the neural dust. For 100 $\mu$m scale sensing nodes embedded 2 mm into the brain, ultrasonic power transmission can enable 7 \% efficiency power links (-11.6 dB), resulting in a received power of $\sim$500 $\mu$W with a 1 mm$^2$ interrogator, which is >10$^7$ more than EM transmission at similar scale (40 pW). Extreme efficiency of ultrasonic transmission and CMOS front-ends can enable the scaling of the sensing nodes down to 10's of $\mu$m.
\end{abstract}


\dropcap{H}alf a century of scientific and engineering effort has yielded a vast body of knowledge about the brain as well as a set of tools for stimulating and recording from neurons across multiple brain structures. However, for clinically relevant applications such as brain-machine interfaces (BMI), a tetherless, high density, chronic interface to the brain remains as one of the grand challenges of the 21st century.

Currently, the majority of neural recording is done through the direct electrical measurement of potential changes near relevant neurons during depolarization events called \emph{action potentials} (AP). While the specifics vary across several prominent technologies, all of these interfaces share several characteristics: a physical, electrical connection between the active area inside the brain and electronic circuits near the periphery (from which, increasingly, data is sent out wirelessly from a "hub") \cite{Biederman, Fan, Miranda, Szuts}; a practical upper bound of several hundred implantable recording sites \cite{Stevenson, Nicolelis, Harrison1, Carmena}; and the development of a biological response around the implanted electrodes which degrades recording performance over time \cite{Turner, Polikov, Chestek, Suner}. To date, chronic clinical neural implants have proved to be successful in the short range (months to a few years) and for a small number of channels (10's to 100's) \cite{Alivisatos2}. Chronic recording from thousands of sites in a clinically relevant manner with little or no tissue response would be a game changer. 

Outside the scope of clinical neuroprosthetics, the need for large scale recording of ensembles of neurons was recently emphasized by the Brain Research through Advancing Innovative Neurotechnologies (BRAIN) initiative in April 2013 by U.S. President Obama \cite{BRAIN} and several related opinion papers \cite{Alivisatos2, Alivisatos1}. Currently, there are numerous modalities with which one can extract information from the brain. Advances in imaging technologies such as functional magnetic resonance imaging (fMRI), electroencephalography (EEG), positron emission tomography (PET), and magnetoencephalography (MEG) have provided a wealth of information about collective behaviors of groups of neurons \cite{Buzsaki}. Numerous efforts are focusing on intra- \cite{Xie} and extra-cellular \cite{Du2} electrophysiological recording and stimulation, molecular recording \cite{Zamft}, optical recording \cite{Ziv}, and hybrid techniques such as opto-genetic stimulation \cite{Cardin} and photo-acoustic \cite{Filonov} methods to perturb and record the individual activity of neurons in large (and, hopefully scalable) ensembles. All modalities, of course, have some fundamental tradeoffs and are usually limited in temporal or spatial resolution, portability, power, invasiveness, etc. Note that a comprehensive recent review of tradeoffs focused on recording from all neurons in a mouse brain can be found in Marblestone et al. \cite{Marblestone}.

\section{System Concept}
\emph{Low-power CMOS circuits coupled with ultrasonic harvesting and backscatter communication can provide a toolset from which to build scalable, chronic extracellular recording systems.}
\\
\\
This paper focuses on the fundamental system design trade-offs and ultimate size, power, and bandwidth scaling limits of systems built from low-power CMOS coupled with ultrasonic power delivery and backscatter communication. In particular, we propose an ultra-miniature as well as extremely compliant system, shown in {\bf Fig. 1}, that enables massive scaling in the number of neural recordings from the brain while providing a path towards truly chronic BMI. These goals are achieved via two fundamental technology innovations: 1) thousands of 10 -- 100 $\mu$m scale, free-floating, independent sensor nodes, or \emph{neural dust}, that detect and report local extracellular electrophysiological data, and 2) a \emph{sub-cranial interrogator} that establishes power and communication links with the neural dust. The interrogator is placed beneath the skull and below the dura mater, to avoid strong attenuation of ultrasound by bone and is powered by an external transceiver via RF power transfer. During operation, the sub-cranial interrogator couples ultrasound energy into the tissue and performs both spatial and frequency discrimination with sufficient bandwidth to interrogate each sensing node. Neural dust can be either an \emph{active node}, which rectifies or recovers power at the sensing node to activate CMOS electronics for data pre-processing, encoding, and transmission, or a \emph{passive node}, which maximizes the reflectivity of the dust as a function of a measured potential. For both schemes, neural dust can communicate the recorded neural data back to the interrogator by modulating the amplitude, frequency, and/or phase of the incoming ultrasound wave. The descriptions of each scheme and the modulation mechanism of each sensing node are detailed in the later sections.
\\
\\
\emph {Several energy modalities exist for powering and communicating with implants, but many of them are unsuitable for the size scales associated with neural dust.}
\\
\\
The requirements for any computational platform interfacing with microelectrodes to acquire useful neural signals (e.g., for high quality \break motor control) are fairly stringent \cite{Harrison1, Muller}. The two primary constraints on the implanted device are size and power. These are discussed in greater detail below, but we list them briefly next. First, implants placed into cortical tissue with scales larger than one or two cell diameters have well-documented tissue responses which are ultimately detrimental to performance and occur on the time-scale of months \cite{Seymour, Marin}. Note that some debate exists as to what role mechanical anchoring outside the cortex plays in performance degradation. Second, all electrical potentials (extra-cellular or otherwise) are by definition measured differentially, so as devices scale down and the distance between recording points decreases accordingly, the absolute magnitude of the measured potential will also decrease. This decreased amplitude necessitates reductions in the front-end noise, which in turns requires higher power (i.e., for a fixed bandwidth, lowering the noise floor requires increased power consumption). Smaller devices, however, collect less power, and building sufficiently low-power electronics may be extremely challenging. Additionally, to eliminate the risk of infection associated with the transcutaneous/trans-cranial wires required for communication and power, such tethers should be avoided as much as possible; a wireless hub is therefore essential to relay the information recorded by the device through the skull.
\\
\\
\emph{High attenuation in brain tissue and geometry-dependent magnetic coupling limit the transfer efficiency of electromagnetics, especially for miniature implants.}
\\
\\
The most popular existing wireless transcutaneous energy transfer technique relies on electromagnetics (EM) as the energy modality \cite{Rabaey}. An external transmitter generates and transfers information through purely electric \cite{Sodagar} or magnetic \cite{Lee} near field or electromagnetic far field coupling \cite{Poon}; this energy can be harvested by the implanted device and converted into a stable DC supply voltage. Energy transmission via magnetic near field has been used in a wide variety of medical applications and is the principal source of power for cochlear implants \cite{Clark}. As EM requires no moving parts or the need for chemical processing or temperature gradients, it is considered more robust and stable than other forms energy scavenging. When used in-body, however, EM coupling power density is restricted by the potential adverse health effects associated with excess tissue heating in the vicinity of the human body due to electromagnetic fields. This is regulated by the well known FCC and IEEE-recommended levels \cite{IEEE}. Roughly, the upper limit for EM power density transiting through tissue is set by the minimum required to heat a model sample of human tissue by 1$^\circ$C. For electromagnetic waves, the output power density is frequency dependent and cannot exceed a maximum of 10 mW/cm$^2$.

\begin{figure}[t]
\begin{center}
\centerline{\includegraphics[width=.48\textwidth]{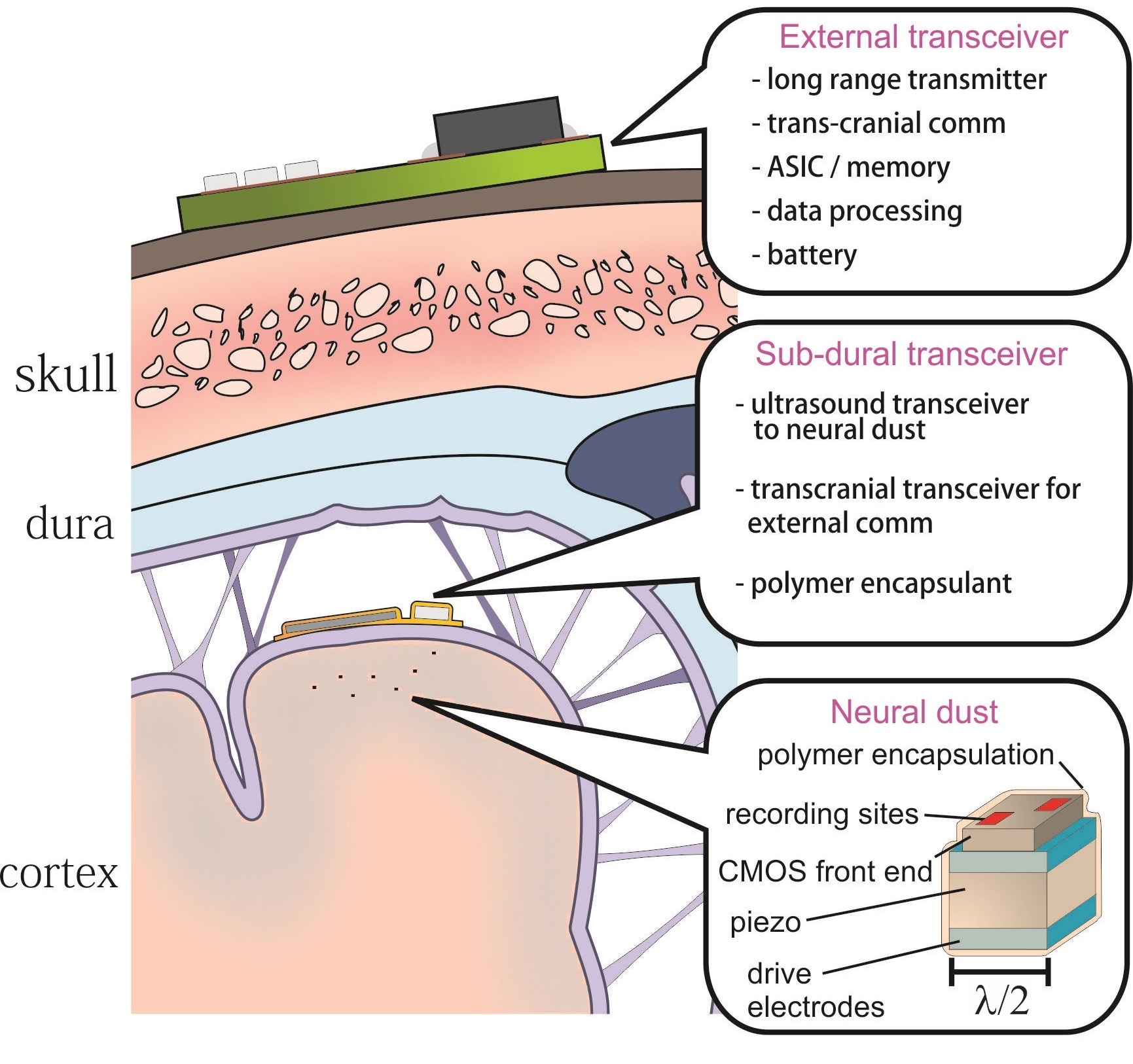}}
\caption{Neural dust system diagram showing the placement of ultrasonic interrogator under the skull and the independent neural dust sensing nodes dispersed throughout the brain.}\label{fig1}
\end{center}
\end{figure}

Consider, in this context, the problem of transmitting EM power to (and information from) very small CMOS chiplets embedded in tissue; does this approach scale to allow high density neural recordings? Regardless of the specific implementation, any such chiplet will contain a resonant component that couples to the EM waves; such a system can be modeled as a series/parallel RLC (for the purposes of this exercise, one may presume that a suitable method exists for modulating the quality factor or mutual coupling of the RLC as a function of neural activity). Given this, the performance of electromagnetic power transfer suffers from two fundamental issues. First, the extreme constraint on the size of the node limits the maximum achievable values of the passives. Assuming a planar square loop inductor, calculations predict the resonant frequency of a 100 $\mu$m neural dust would be $\sim$10 GHz. {\bf Fig. 2 (a)} plots the modeled channel loss, or the attenuation of the EM signal as it propagates through 2 mm of brain tissue, due to tissue absorption and beam spreading, as a function of frequency. We observe that there is an exponential relationship between the channel loss and the frequency, and at 10 GHz -- the total combined loss for one-way transmission is approximately 20 dB. Moreover, at these very small footprints (compared to the wavelength, which is in millimeter range), the receive antenna efficiency becomes quite small, thereby easily adding roughly 20 dB of additional loss, resulting in a total gain of at most -40 dB. The tissue absorption loss penalty incurred by operating at a high frequency can be reduced by increasing the capacitance density using 3D inter-digitized capacitor layouts, but even then, as shown in {\bf Fig. 2 (b)}, scaling down the dimensions of the chiplets increases the resonant frequency of the link, causing an exponential increase in the tissue absorption loss and the overall channel loss, and the efficiency of EM transmission becomes miniscule.

\begin{figure}[t]
\begin{center}
\centerline{\includegraphics[width=.485\textwidth]{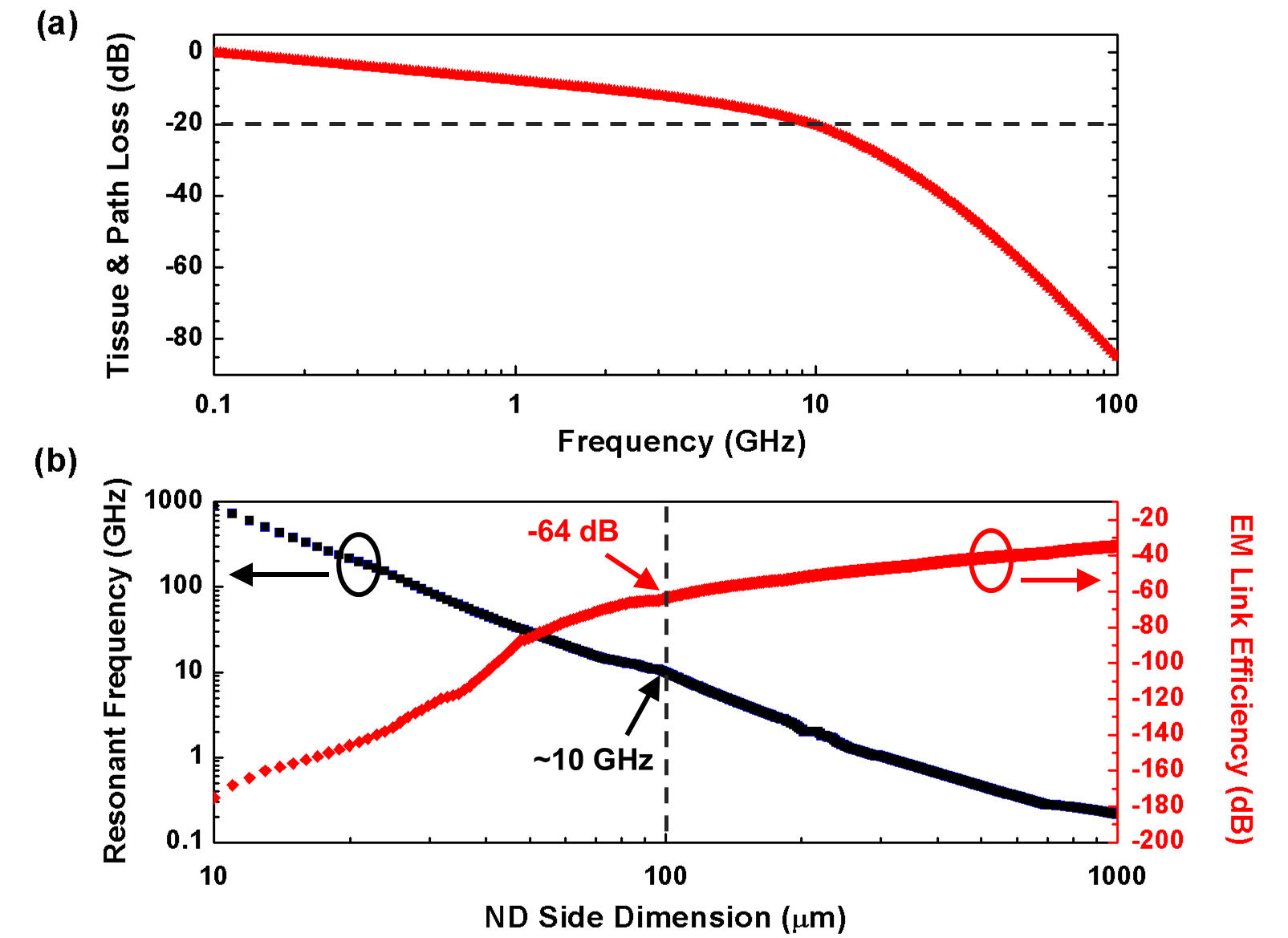}}
\caption{{\bf (a)} Total channel loss in 2 mm brain tissue, due to both tissue and propagation loss, increases exponentially with frequency, resulting in a 20 dB of loss at 10 GHz. {\bf (b)} The mutual coupling, and therefore link efficiency, also reduces dramatically with the scaling of chiplet dimensions.}
\label{fig2}
\end{center}
\end{figure}

To make matters worse, the mutual coupling between the transmitter and receiver coils drops dramatically and significantly degrades the transfer efficiency and increases the sensitivity to misalignments \cite{Salim, Fotopoulou}. As shown in {\bf Fig. 2 (b)}, EM transmission with a 100 $\mu$m neural dust embedded 2 mm into the cortex results in 64 dB of transmission loss. Given a 1 mm$^2$ transmitter aperture outputting 100 $\mu$W of power -- limited by the need to satisfy safety regulations on output power density of 10 mW/cm$^2$ -- the resulting received power at the neural dust is $\sim$40 pW. This is orders of magnitude smaller than the power consumption imposed by noise requirements on the front-end amplification circuitry (refer to later sections for further discussion). As a result, prior work by \cite{Biederman}, which features the most energy-efficient and smallest wirelessly EM powered neural recording system to date, at 2.5 $\mu$W/channel and 250 $\mu$m x 450 $\mu$m, is limited in terms of further dimensional scaling and increasing the range (the effective range within brain tissue for this work was 0.6 mm). We conclude that due to the non-linear interplay of form factor, speed of light, and frequency spectra of tissue absorption, EM power transmission is not an appropriate energy modality for the powering of 10's of $\mu$m sized neural dust implants. 
\\
\\
\emph{Ultrasound is attractive for in-tissue communication given its short wavelength and low attenuation.}
\\
\\
Ultrasonic transducers have found application in various disciplines including imaging, high intensity focused ultrasound (HIFU), nondestructive testing of materials, communication and power delivery through steel walls, underwater communications, transcutaneous power delivery, and energy harvesting \cite{Ishida, Wong, Ozeri, Richards}. The idea of using acoustic waves to transmit energy was first proposed in 1958 by Rosen \cite{Rosen} to describe the energy coupling between two piezoelectric transducers. Unlike electromagnetics, using ultrasound as an energy transmission modality never entered into widespread consumer application and was often overlooked because the efficiency of electromagnetics for short distances and large apertures is superior. However, at the scales discussed here and in tissue (i.e., aqueous media) the low acoustic velocity allows operation at dramatically lower frequencies, and more importantly, the acoustic loss in tissue is generally substantially smaller than the attenuation of electromagnetics in tissue ({\bf Table 1}). 

As mentioned earlier, the relatively low acoustic velocity of ultrasound results in substantially reduced wavelength compared to EM. For example, 10 MHz ultrasound in brain tissue has a wavelength $\lambda$ = 150 $\mu$m, while for 10 GHz EM, $\lambda$ = 5 mm \cite{Hoskins}. This smaller wavelength implies that for the same transmission distance, ultrasonic systems are much more likely to operate in the far-field, and hence offer more isotropic characteristics than an EM transmitter (i.e., the ultrasonic radiator can obtain larger spatial coverage). This opens up the prospect of interrogation of multiple nodes via frequency binning. More importantly, the acoustic loss in brain tissue is fundamentally smaller than the attenuation of electromagnetics in tissue because acoustic transmission relies on compression and rarefaction of the tissue rather than time-varying electric/magnetic fields that generate displacement currents on the surface of the tissue \cite{Leighton}. This is also manifested by the stark difference in the time-averaged acceptable intensity for ultrasound for cephalic applications, regulated by FDA, which is approximately 9x (94 mW/cm$^2$) for general-purpose devices and 72x (720 mW/cm$^2$) more than EM for devices conforming to output display standards (ODS) (recall EM is limited to 10 mW/cm$^2$) \cite{FDA}. 

As an aside, in order to increase the instantaneous power captured by an implant, FDA regulations would allow an interrogator to transmit up to 190 W/cm$^2$ of spatial peak pulse-averaged power density. This approach, however, must be taken with caution as more in-depth studies of the thermal impact of duty-cycled operation on the tissue are necessary to determine safe parameters of the applied duty-cycle and meet the time-averaged power level constraint \cite{Tufail, King}. Also, as demonstrated by a body of work investigating the effectiveness of ultrasound as a means of modulating neuronal activity \cite{Foley, Krasovitski, Tyler, Hameroff}, systems operating in this regime may be capable of micro-stimulating the brain at a CW time-averaged output intensity as low as 1 W/cm$^2$ \cite{Tsui}, and cause tissue ablation through heating and cavitation at intensities in the focal region of 100 - 1000 W/cm$^2$ \cite{Zhou}.
\\
\\
\emph{Piezoelectric ultrasonic transducers suitable for implanted applications are available.}
\\
\\
Piezoelectricity refers to the phenomenon present in certain solid (usually crystalline) materials where there is an interaction between the mechanical and electrical states. As a result, piezoelectric materials can transduce electrical energy into mechanical energy and vice versa by changing lattice structure, and this state change is accessible via either electrical stimulation or mechanical deformation. These materials serve as a critical component in the construction of probes that generate ultrasonic waves to enable ultrasound technology used in the medical industry. A relatively wide range of piezoelectric materials are available, each suitable for different applications. For instance, materials such as single crystal lithium niobate (LiNbO$_3$) and polymer PVDF are excellent choices for fabricating large aperture single element transducers due to their low dielectric permittivity \cite{Shung}. On the other hand, a ceramic compound known as lead zirconate titanate (PZT) is a popular choice for high performance diagnostic ultrasonic imaging due to its greater sensitivity, higher operational temperature, and exceptional electromechanical coupling coefficient. The electromechanical coupling coefficient is a figure of merit used to describe the ability of a material to convert one form of energy into another, and is defined as the ratio of stored mechanical energy to total stored energy in a given material. The lead content of PZT makes it difficult to introduce into human tissue in chronic applications; several works have demonstrated encapsulation as an option to avoid this issue \cite{Zenner, Maleki}, but the long-term stability of such encapsulation layers remain to be investigated. 
\begin{figure}[t]
\begin{center}
\centerline{\includegraphics[width=.5\textwidth]{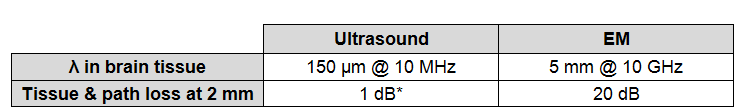}}
\end{center}
{\dospecialaccents\xfigtextfont {\bf \small Table 1.} \small Comparison of both the scale and the loss incurred in brain tissue between ultrasound and EM radiation, displaying the stark differences in the achievable spatial resolution (set by the wavelength) and the tissue/path loss for operating frequency of a 100 $\mu$m neural dust (*Attenuation of ultrasound in brain is 0.5 dB/(cm$\cdot$MHz) \cite{Hoskins})}.
\end{figure}

Luckily, biocompatible piezoelectric materials exist with properties similar (but generally inferior) to PZT; these include barium titanate (BaTiO$_3$), aluminum nitride (AlN) and zinc oxide (ZnO) \cite{Przybyla}. Although the dielectric coefficients of AlN and ZnO are less than one-hundredth that of BaTiO$_3$ (which can result in an improvement in the signal to noise ratio due to the lower parallel plate capacitance), their piezoelectric coefficient (which is critical to the link efficiency) is one-tenth that of BaTiO$_3$. Therefore, BaTiO$_3$ transducers are assumed for the remainder of the paper. Clearly, material engineering to synthesize higher performance piezoelectric composite materials and reliability studies to assess performance over extended periods of operation are both active areas of research that can significantly contribute to the realization of neural dust.

\section{System design and constraints: Power Delivery}

There are several implementation strategies for the neural dust. A neural dust can be an \emph{active node}, which consists of a piezoelectric transducer to recover power at the sensing site to activate CMOS electronics for data pre-processing, encoding, and transmission, or a \emph{passive node}, which maximizes the reflectivity of the dust as a function of a measured potential. In an active node scheme, the design of neural dust is heavily constrained in both size and available power to the implant. As a result, it is imperative to accurately model the transmission channel to maximize the power efficiency. Therefore, this section elaborates design tradeoffs and methodologies for power delivery optimization.
\\
\\
\emph{The propagation characteristics of ultrasound must be considered in determining the maximum range of neural dust and the optimal dimension of the external interrogator.}
\\
\\
As the pressure field generated by a uniform continuous-wave excited piezoelectric transducer propagates through the tissue medium, the characteristics of the pressure field change with distance from the source. The varying field is typically divided into two segments, \emph{near field} and \emph{far field}. In the near field, the shape of the pressure field is cylindrical and the envelope of the field oscillates. At some point distal to the transducer, however, the beam begins to diverge and the pressure field becomes a spherically spreading wave, which decays inversely with distance. The transition between the near and far field is where the pressure field converges to a natural focus, and the distance at which this occurs is called the Rayleigh distance, defined as,

\begin{equation}
L=\mfrac{(D^2 - \lambda^2)}{4\lambda} \approx \mfrac{D^2}{4\lambda} , D^2\gg \lambda^2
\end{equation}
where D is the aperture width of the transmitter and $\lambda$ is the wavelength of ultrasound in the propagation medium. In order to maximize the received power, it is preferable to place the receiver at one Rayleigh distance where the beam spreading is at a minimum. Therefore, with 2 mm of transmission distance and a resonant frequency of 10 MHz ($\lambda$ = 150 $\mu$m), the maximum dimension of the external interrogator should be $\sim$1 mm.
\\
\\
\emph{Neural dust transducers can be simulated with finite element packages and incorporated into a KLM-based link model.}
\\
\\
Due to the importance of piezoelectric transducers in various applications, a number of models of the electromechanical operation of one-dimensional piezoelectric and acoustic phenomena have evolved over the years. The KLM model is arguably the most common equivalent circuit and is a useful starting point to construct a full link model with the intent of examining scaling and system constraints \cite{KLM}. The basic model includes a piezoelectric transducer with electrodes fully covering the two largest faces of the transducer. The entire transducer is modeled as a frequency-dependent three-port network, consisting of one electrical port (where electric power is applied or collected) and two acoustical ports (where mechanical waves are produced or sensed from the front and back faces of the transducer). The parallel-plate capacitance due to the electrodes and the frequency-dependent acoustic capacitance are modeled as C and $X_i$, respectively, and the transduction between electrical and mechanical domains is modeled as an ideal electromechanical transformer with a turn ratio of $\Phi$, connected to the middle of a transmission line of length $\lambda$/2, as shown in {\bf Fig. 3}. Assuming an infinite 2D plate piezoelectric transducer of thickness \emph{h}, the resonant frequency is set by \emph{h} = $\lambda$/2; at the resonant frequency, the ultrasound wave impinging on either the front or back face of the transducer will undergo a 180$^{\circ}$ phase shift to reach the other side, causing the largest displacement between the two faces. This observation implies that phase inversion only exists at the odd harmonics of the fundamental mode in a given geometry.

The KLM model, however, was derived under the assumption of pure one-dimensional thickness vibration, and therefore can only provide a valid representation for a piezoelectric transducer with an aspect ratio (width/thickness) greater than 10 that mainly resonates in the thickness mode \cite{Roa-Prada}. Given the extreme miniaturization target for the neural dust, a cube dimension (aspect ratio of 1:1:1) is a better approximation of the geometry than a plate (aspect ratio > 10:10:1). Due to Poisson's ratio and the associated mode coupling between resonant modes along each of the three axes of the cube, changing aspect ratio alters the resonant frequencies \cite{Holland}. The piezoelectric transducers for both the interrogator and the neural dust must be designed to resonate at the same frequency to maximize the link efficiency. In the model below, we assume the neural dust nodes are cubic and the external transceiver is approximately planar (i.e., 2D) so care must be taken not to confuse the thickness of the interrogator and the neural dust.

In order to obtain KLM parameters for the neural dust transducer, we simulated a cube transducer using a 3D finite element package (COMSOL Multiphysics) to model both the resonant frequency shift vs. a plate and the manifestation of spurious tones and higher resonances. The effect of resonance shift is included in the KLM model by extracting the effective acoustic impedance of the neural dust from the COMSOL model. To match the resonant frequency of the interrogator and the neural dust, the interrogator thickness is varied to match the fundamental thickness mode of the neural dust. Approximately 66 \% of the total output energy is contained in the main thickness resonance; this is modeled as a loss term. Coupling into other modes, however, can be reduced by stretching BaTiO$_3$ in the [110] direction because BaTiO$_3$ is both anisotropic and partially auxetic, exhibiting negative Poisson's ratio and therefore providing gain when stretched \cite{Baughman, Aleshin}. Well-engineered placement of electrodes may enable orientation-insensitive implant nodes and can allow multi-node ad-hoc type communication networks. More on this topic will be elaborated in the discussion section.
\\
\\
\emph{The maximum energy transfer efficiency can be found via a link model consisting of a cascade of two-port networks.}
\\
\\
A good model of the ultrasonic channel is crucial in order to assess the tradeoffs in optimizing systems for energy transfer through lossy brain tissue. The complete energy link model is shown in {\bf Fig. 4 }and can be divided into three parts: (1) the ultrasonic interrogator or \emph{transmitter}, (2) tissue, and (3) the neural dust or \emph{receiver}. A signal generator and amplifying stages produce power for the ultrasonic transmitter through an impedance matching circuit that provides conjugate matching at the input. The ultrasonic wave launched by the interrogator penetrates brain tissue, modeled as a lossy transmission line, and a fraction of that energy is harvested by the ultrasonic receiver, or neural dust. We evaluated embedding the receiver up to 2 mm into the tissue, which generates an AC voltage at the electrical port of the piezoelectric transducer in response to the incoming ultrasonic energy.

\begin{figure}[t]
\begin{center}
\centerline{\includegraphics[width=.5\textwidth]{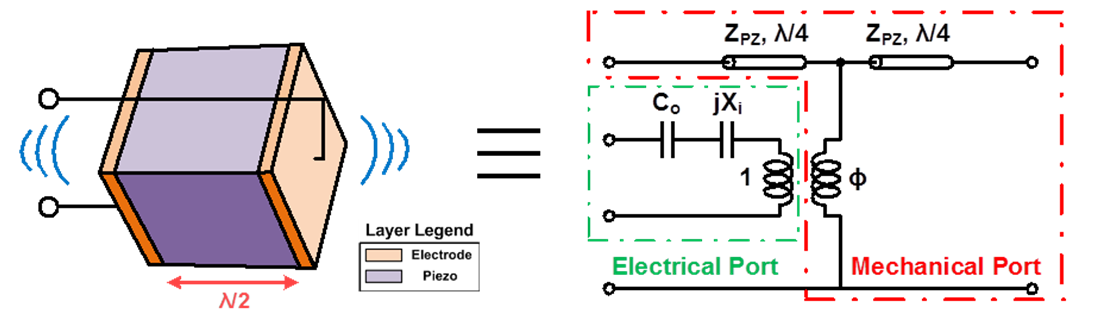}}
\caption{KLM model of a neural dust piezoelectric transducer, showing one electrical port and two mechanical ports. Coupling between the domains is modeled with an ideal electromechanical transformer.}\label{fig3}
\end{center}
\end{figure}

In order to compute the link energy transfer efficiency, the model can be decomposed to a set of linear and time-invariant two-port parameters, representing a linear relationship between the input and output voltage. Here, we choose to represent the input-to-output relationship using ABCD parameters, which simplify analysis of cascades of two-port networks through simple matrix multiplication. By representing the link model with the two-port network, we can come to conclusions concerning optimal power transfer efficiency (or "gain"). 

Generally, maximum link efficiency ($G_{max}$) is achieved when we can conjugate match both the input and the output of a two-port network. However, with a 100 $\mu$m neural dust node, the output impedance level is such that it would require $\sim$100 $\mu$H of inductance to perfectly conjugate match the output of the two port link network. Given the compact form factor of the neural dust, it is completely infeasible to obtain such inductance with electrical means, and therefore $G_{max}$ is an unachievable figure of merit. It may be possible to approach $G_{max}$ by mechanical means such as the addition of material layers that perform an acoustic impedance transformation, or similarly, by electromechanical means such as utilizing micromachined acoustic resonators. We do not explore the first option in detail as it would likely lead to thickness increases on order of integer fractions of a wavelength (but see {\bf Fig. 5 (b)} and below); the second option is touched upon in Discussion and Conclusion. Therefore, for comparison and scaling analysis, we assume we only have impedance control at the input, or the interrogator side, and therefore, power gain ($G_p$) is the suitable figure-of-merit.
\\
\\
\emph{For a 100 $\mu$m node embedded 2 mm into the brain, ultrasonic power transmission can enable 7 \% efficiency power links (-11.6 dB), resulting in a received power of $\sim$500 $\mu$W with a 1 mm$^2$ interrogator.}
\\
\\
The complete link model is implemented in MATLAB with the limitations of the KLM model (as outlined in the previous section) corrected via COMSOL simulations. Given a 1 mm$^2$ interrogator, {\bf Fig. 5} plots both the efficiency of the link and the received power at the sensor node as the size of the dust node scales and the thickness of the interrogating transducer is adjusted to match the resonant frequency of the dust node and the tissue (i.e., transmission line resonator). We note that the maximum efficiency of the KLM-adapted link model, where the interrogator is fully immersed in the tissue medium, is limited to 50 \% because both the back and front side of the interrogator are loaded by the tissue layer. This results in an efficiency drop of 3 dB as the ultrasonic energy couples to both the front and back face of the transducer equally. Additionally, without any impedance matching, since the acoustic impedance of the tissue (1.5 MRayls) and that of BaTiO$_3$ (30 MRayls) are drastically different, significant reflection occurs at their boundaries. Depending on the thickness of neural dust and the resonant frequency of the network, ultrasonic waves launched by the interrogator undergo varying phase changes through the lossy tissue. Thus, the efficiency of a system with smaller dust nodes can be improved if the total propagation distance happens to be a multiple of a wavelength of the ultrasound. As a result, for dust nodes greater than 100 $\mu$m, we note that the efficiency does not monotonically increase with the dimension. On the other hand, for a dust node that is less than 100 $\mu$m in dimension, because the wavelength associated with the network's resonant frequency is much smaller than its tissue propagation distance, the link efficiency depends more heavily on the cross-sectional area of the neural dust. Therefore, we note that the efficiency will drop at least quadratically with the reduction of neural dust dimension. The efficiency of the link can be improved with a $\lambda$/4 matching layer for impedance transformation, but the improvement is limited due to the loss from the material (e.g., attenuation of graphite epoxy is $\sim$16 dB/(cm$\cdot$MHz) \cite{Mills} compared to that in brain tissue which is 0.5 dB/(cm$\cdot$MHz) \cite{Hoskins}) as shown in {\bf Fig. 5 (b)}. Note that for the case with this matching layer, the efficiency is worse for dust nodes that are >500 $\mu$m since the loss of the matching layer outweighs that of the tissue.

\begin{figure}[b]
\begin{center}
\centerline{\includegraphics[width=.47\textwidth]{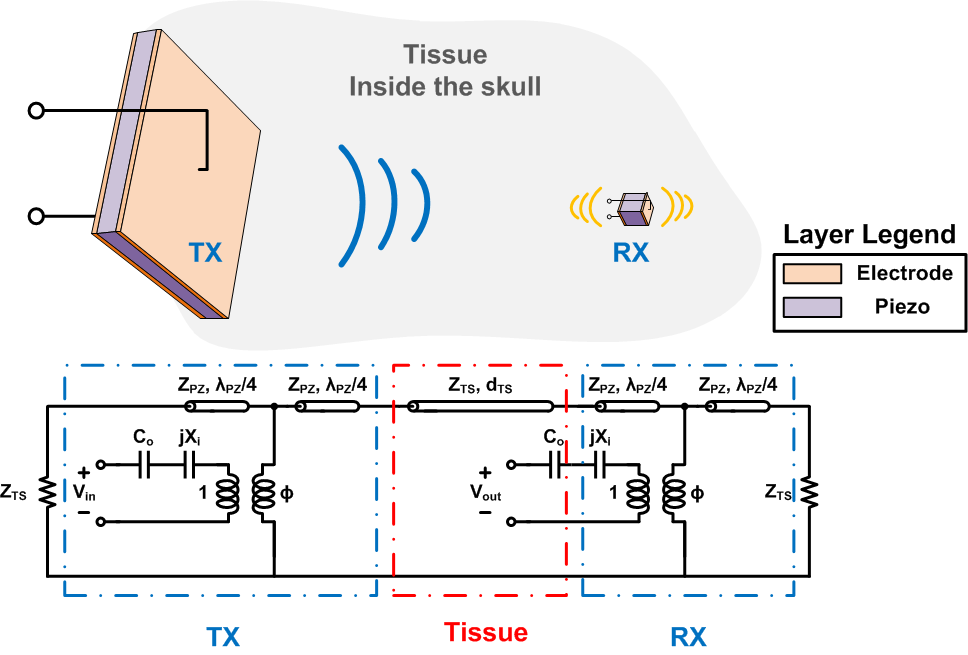}}
\caption{Complete single interrogator, single neural dust power and communication through link models.}\label{fig4}
\end{center}
\end{figure}

More specifically, simulation of the complete link indicates that for a 100 $\mu$m node embedded 2 mm into the brain, ultrasonic power transmission can enable 7 \% efficiency power transmission (-11.6 dB). As shown in {\bf Fig. 5 (a)}, the optimal transmission frequency is 8 MHz; half of this peak $G_p$ can be maintained for carrier frequencies that are $\pm$2 MHz separated from this peak. At the resonant frequency, we can receive up to $\sim$500 $\mu$W at the neural dust node (resulting in nano-meters of total displacement) with a 1 mm$^2$ interrogator, which is >10$^7$ more than EM transmission at the same size scale (40 pW in {\bf Fig. 2}). Scaling of neural dust also indicates that approximately 3.5 $\mu$W can be recovered by a dust node as small as 20 $\mu$m through ultrasonic transmission, which is still in the realm of feasibility to operate a state-of-the-art CMOS neural front-end. Designing an ultra-energy efficient neural front-end in CMOS in such small footprint (20 $\mu$m x 20 $\mu$m), however, is an extremely challenging problem and is discussed in detail below.

\section{System design and constraints: Sensing / Communication}
\emph{Extracting neural potential recording from a noisy environment is a challenging problem.}
\\
\\
The electrical activity of neurons is most directly measured as an electrical potential across the cellular membrane. As a result, the highest fidelity measurement can be achieved using patch-clamp methods, where a glass pipette is placed in the vicinity of the cell and an intra-cellular electrical connection is established by penetrating the cellular membrane and sealing the membrane around the pipette. While this approach is well studied and commonly practiced, it does not scale well and is currently not useful for chronic implants due to the complexity of the procedure (but see \cite{Kodan, Robinson, Yao}). Due to these limitations, clinically-relevant, implantable recordings are taken \emph{extra-cellularly}; that is, electrical measurements are taken entirely outside the cells.

\begin{figure}[b]
\begin{center}
\centerline{\includegraphics[width=.5\textwidth]{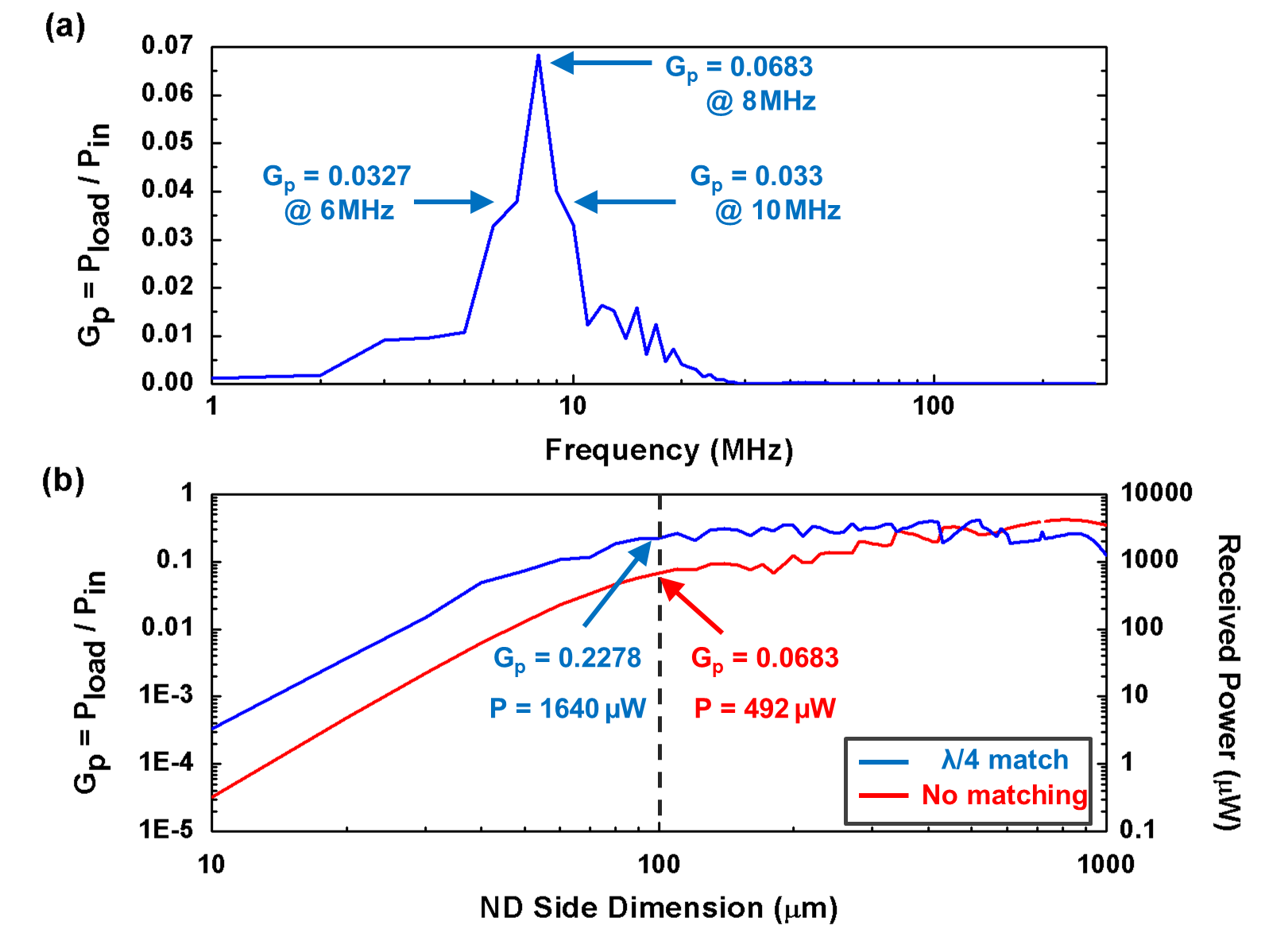}}
\caption{{\bf (a)} Ultrasonic power transfer efficiency vs. operating frequency for a 100 $\mu$m neural dust {\bf (b)} Link efficiency with and without a matching layer as a function of the neural dust side dimension.}\label{fig5}
\end{center}
\end{figure}

A typical extracellular electrophysiological recording of neural activity in tissue usually records electrical potential differences between one electrode placed in-tissue near the neural activity and a second electrode "far away" which acts as a global ground or counter electrode (depending on the configuration). The recorded signal consists of three components: an electrochemical offset that appears as a DC offset, typically in the range of 100's of mV, low-frequency (0.1 -- 600 Hz) changes \cite{Belitski} ($\sim$0.5 mV amplitude) often termed \emph{local field potential} (LFP) from a spatial average of neural activity in the neighborhood of electrodes and high frequency (0.8 -- 10 kHz) \emph{action potential} (AP) or spiking events ($\sim$100 $\mu$V) associated with the discharge of individual neurons in the vicinity of the electrode \cite{Nicolelis}. Ignoring noise inherent in the recording equipment (which is usually not insubstantial), there are two main sources of cortical recording noise: thermal noise generated by the recording electrode and the tissue interface and biological interference which arises from asynchronous neural activity in close proximity to the recording site. Therefore, neural signal acquisition chains often rely on obtaining a maximum signal level at the front-ends and/or separating the $\mu$V-level desired signal from large offsets and low frequency disturbances.
\\
\\
\emph{Spatial separation of recording electrodes to maximize the achievable differential signal on neural dust is the bottleneck for scaling.}
\\
\\
Free-floating extracellular recording at untethered, ultra-small dust nodes poses a major challenge in scaling. Unlike the needle-like microelectrode shanks that can measure time-domain electrical potential at each recording site in relation to a common electrode, placed relatively far away, both the recording and the common electrode must be placed within the same (very small) footprint. Although the two are interchangeable, the separation and therefore, the maximum differential signal between the electrodes are inherently limited by the neural dust footprint, and follow the dipole-dipole voltage characteristic that decreases quadratically (unless very near a cell body, in which case it appears to scale exponentially; see \cite{Gold} for a more thorough review) with increasing separation distance. Since the power available to the implant has a fixed upper bound (see above), the reduction of extracellular potential amplitude as the neural dust dimensions are scaled down in the presence of biological, thermal, electronic, and mechanical noise (which do not scale), causes the signal-to-noise (SNR) ratio to degrade significantly; this places heavy constraints on the CMOS front-ends for processing and extracting the signal from extremely noisy measurements. Therefore, if we consider sufficient SNR at the input of the neural front-ends as one of the design variables, the scaling of neural dust (as depicted in {\bf Fig. 5 (b)}) must be revisited. 
\\
\\
\emph{Careful co-optimization of piezoelectric transducer and CMOS front-end circuitry can push the operation of neural dust down at least to the 50 $\mu$m scale.}
\\
\\
Focusing specifically on the scaling of a cubic neural dust, we run into the inherent limitation in the maximum achievable differential signal (discussed above). At a separation distance of 100 $\mu$m between recording electrodes, we expect a 10 $\mu$V AP amplitude [data derived from \cite{Du}], with the amplitude further reducing quadratically as the separation is reduced. Since the power available to the neural dust is limited, the design goal of a front-end architecture is to minimize the input-referred noise within this power budget. The power efficiency factor (NEF$^2\cdot V_{dd}$) quantifies the tradeoff between power and noise \cite{Muller} and extrapolating from the measurement result of a previous\break CMOS neural front-end design (NEF$^2\cdot V_{dd}$ of 9.42 \cite{Biederman}), we can estimate the relationship between the input-referred noise level and the DC power consumption of an optimally designed front-end architecture as we scale. The fundamental limit to the NEF$^2\cdot V_{dd}$ occurs at a supply voltage of at least $\sim$4 $k_BT/q$ or 100 mV, in order to reliably operate the FET, and by definition, the NEF of 1 for a single BJT amplifier \cite{Steyaert}. In principle, one could push the supply voltage down to $\sim$2 $k_BT/q$, but in practice 100 mV is already extremely aggressive.

Fixing the input SNR to 3, which should be sufficient for extracting neural signals, we can evaluate the scaling capability of neural dust as shown in {\bf Fig. 6}. We assumed the use of BaTiO$_3$ in the model described in the section above and do not include the use of matching layers. We also assumed that the interrogator's output power is constrained by the two different FDA-approved ultrasonic energy transfer protocols. We note that there exists an inherent tradeoff between the power available to the implant and the exponential increase in the power required to achieve an SNR of 3 with the reduction of spacing between the electrodes. The point of intersection in {\bf Fig. 6} denotes the minimum size of neural dust that enables the operation of the complete link. For the stated assumptions, this occurs at 50 $\mu$m, which is greater than the dimension at which the thermal noise from the electrode (R = 20 k$\Omega$ and BW = 10 kHz) limits further scaling. This effectively means that, staying within FDA-approved ultrasound power limits, assuming an SNR of 3 is required, neural dust nodes smaller than 50 $\mu$m cannot receive enough power to distinguish neural activity from noise. Note that the cross-over assumes 100 \% efficiency in the rectifier and zero overhead cost in the remaining circuitry, both of which will not be true in practice (i.e., the actual size limit will be larger than this).

Given the lower size limit for scaling these systems, as well as the need to implant them entirely in the cortex, both wireless power and communication schemes are required for the neural dust nodes. The communication strategy is detailed below.
\\
\\
\emph{Neural electrophysiological data can be reported back via backscattering -- i.e., modulating reflection of the incident carrier.}
\\
\\
Radio frequency identification (RFID) technologies have found broad adoption in the past decade, and were made possible by advances in wireless powering techniques as well as the improved energy-efficiency of the computational substrates. In general, RFID employs two different mechanisms to communicate with sensor tags: active and passive \cite{Weinstein}. When queried, \emph{active tags}, which are battery-powered and contain a low power radio like conventional wireless devices, internally generate electromagnetic radiation in order to transmit the data back to the reader. In contrast, \emph{passive} and \emph{semi-passive} tags transmit data by modulating the incoming RF energy and re-radiating the modulated RF energy back to the reader, a method called \emph{backscattering}. Modulation of the backscattered RF energy can be achieved by varying the load impedance, which changes the coefficient of reflectivity. Furthermore, backscattering is amenable to parallel communication among sensor tags and one interrogator distinguishing among different receivers by using frequency diversity \cite{Finkenzeller}. Multi-mode strategies are discussed in Discussion and Conclusion.

\begin{figure}[t]
\begin{center}
\centerline{\includegraphics[width=.47\textwidth]{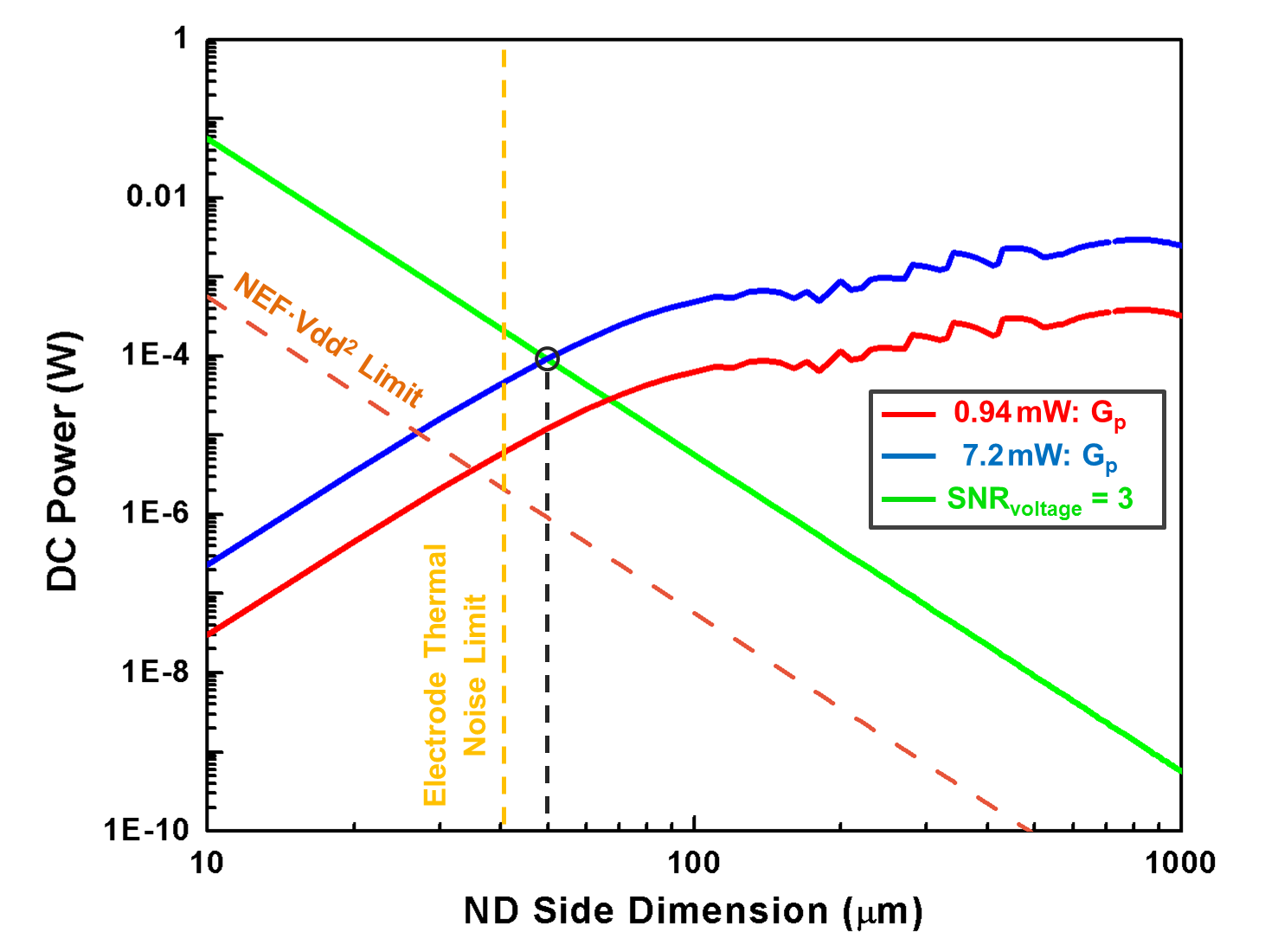}}
\caption{As we scale down the neural dust size, more power is needed to keep the noise floor down to maintain SNR while less power is captured. The intersection of these two trends is the smallest node that will still operate. Scaling with an SNR of 3 shows operation down to 50 $\mu$m. The analysis assumes the use of BaTiO$_3$, two different FDA-approved ultrasonic energy transfer protocols, and does not include the use of matching layers.}
\label{fig7}
\end{center}
\end{figure}

For the ultra-miniature, chronic implants discussed here (which have stringent requirements on both the size and power available to the implant), broadcasting the information back to the interrogator via backscattering is a more attractive choice than building a fully active transmitter on the implant. As a passive device, backscattering receivers do not need batteries or significant capacitive energy storage, thus extending lifetimes, eliminating the risk of battery leakage, and removing the significant impediment to size scaling that would be created by the dramatically reduced capacitance available on a small node. The powering and communication strategies developed for electromagnetic backscattering can be applied to any link, regardless of the transmission channel modality (i.e., ultrasound).
\\
\\
\emph{Co-integration of CMOS and piezoelectric transducer is challenging, but CMOS can provide dynamic control over the load impedance.}
\\
\\
The CMOS component of an active neural dust node must at least consist of a full-wave bridge rectifier to convert the harvested piezoelectric AC signal to a DC level and regulators to generate a stable and appropriate DC supply voltage for the rest of the CMOS circuitry. The basic architecture of the CMOS front-ends will depend on the application. For the acquisition of the entire neural signal trace, we must capture both the LFP and action potentials. Given the relative amplitude, DC offset, and frequency range of these signals, the circuit must operate at a full bandwidth of 0 to 10 kHz with >70 dB of input dynamic range \cite{Muller}. Researchers have demonstrated a mixed-signal data acquisition architecture solution to extract LFP and action potentials, originally proposed in \cite{Muller}, which cancels the DC offset in the analog domain to alleviate the dynamic range constraints and to eliminate bulky passive components used in \cite{Yazicioglu1, Yazicioglu2}. Therefore, the CMOS front-ends include rectifiers, voltage regulators, low-noise amplifiers, DC-coupled analog-to-digital converters (ADC) and modulators to communicate the decoded information back to the interrogator.

Co-integration and packaging challenges and -- most importantly -- the footprint of current CMOS neural front-ends present major roadblocks to the active implant approach. The smallest CMOS neural front-end system published to date, not including rectifiers and modulators, occupies approximately 100 $\mu$m$^2$ of silicon real estate \cite{Muller}, and packing the same functionality onto a smaller footprint may not be plausible. Thinned, multi-substrate integration to meet the volume requirements while keeping the overall CMOS area constant may resolve this issue, but requires substantial further technology development to represent a viable solution. Scaling the active electronics to appropriate dimensions is clearly a bottleneck, but presents an enticing opportunity for further innovation to address the issue.

\section{System design and constraints: Passive node}
\emph{A MOSFET (Metal-Oxide-Semiconductor field effect transistor) may be used to modulate the impedance of the transducer as a function of neural signals, obviating the need for active front-ends.}
\\
\\
Ideally, the simplest neural dust would consist of a piezoelectric transducer with a set of surface electrodes that can record the occurrence of a neural spike, and the extracted measurement can be reported back to the interrogator by somehow encoding the information on top of the incoming ultrasound wave. The design methodology we adopt here is that of elimination: starting with current neural front-end architectures that consist of, but are not limited to, rectifiers, high resolution ADC, amplifiers, regulators and modulators, we start eliminating each component to truly understand its impact on overall system performance, and therefore assess its necessity for inclusion on the dust node itself. Rectifiers and voltage regulators are essential to provide a stable DC power supply for the transistors in the system. In order to prevent variations in the electrical response of the circuits with the variation of its power supply, it is important to have sufficient amount of capacitance to curb any supply ripple and filter out high frequency electrical noise. As a result, these two components tend to occupy the largest amount of space in the CMOS die footprint. 

Here, let us re-examine the need for a DC supply as we entertain the idea of completely eliminating both the rectifiers and the voltage regulators. In this scenario, the piezoelectric transducer harvests the incoming ultrasonic wave and directly converts it to an AC electrical voltage. At this point, the design goal essentially boils down to devising ways of encoding neural data on top of this incoming ultrasound wave, to be reported back to the interrogator via modulation. 

We propose a method outlined in {\bf Fig. 7}, where the drain (D) and source (S) of a single FET sensor are connected to the two terminals of a piezoelectric transducer while the FET modulates the current $I_{DS}$ as a function of a gate (G) to source voltage, $V_{GS}$. In this scheme, given that the supplied $V_{DS}$ of the FET is an AC voltage that swings both positive and negative, the body (B) of the FET must be biased carefully. Normally, for an NFET, the body is connected to the source voltage to prevent the diode at the B-S and B-D junctions from turning on. However, keep in mind that since a FET is a symmetric device, the source and drain are defined only by which terminal is at a lower potential. Therefore, the electrical source/drain terminals, or left/right for disambiguation (from a cross section of MOS device), swap physical sides every half cycle of the harvested AC waveform. As a result, simply shorting the body to either physical terminal of the FET causes the diode formed at the B-S and B-D junctions to be forward-biased, so care must be taken to avoid neural signal from modulating the incoming sinusoid only half of the cycle. 

As a result, we propose an alternative biasing scheme for the FET to modulate the entire sinusoid as shown in {\bf Fig. 7}. The resistors $R_b$ act to cause the neural potential to appear between the gate and both of the left/right terminals of the transistors while superimposing the AC waveform from the ultrasonic transducer across these same two terminals. In this manner, even though the electrical source/drain terminals swap every half cycle, during both halves of the cycle the $V_{GS}$ of the FET is modulated by the neural signal.

\begin{figure*}[b]
\begin{center}
\centerline{\includegraphics[width=.885\textwidth]{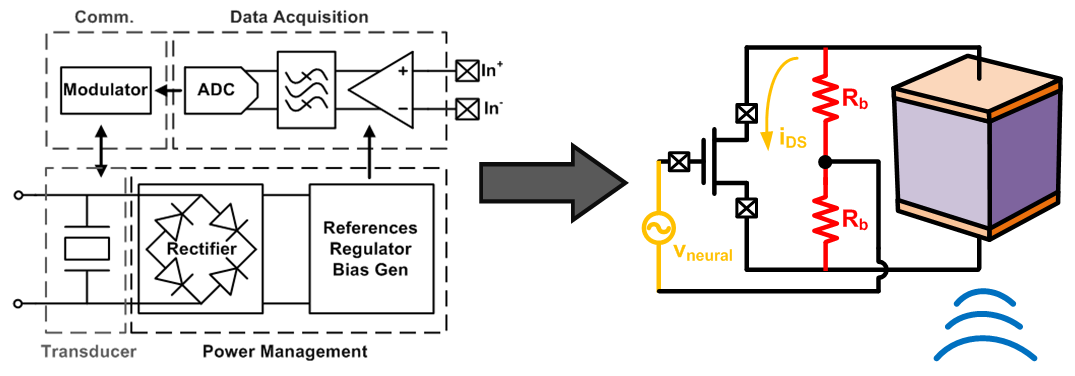}}
\caption{A process of elimination leads to a simple architecture (right) where we utilize a FET to vary the electrical load impedance, changing the ultrasonic wave reflectivity at the dust and modifying the backscattered wave.}
\label{fig8}
\end{center}
\end{figure*}

The circuit achieves this superposition by relying on the fact that the neural signals occupy a much lower frequency band than the ultrasound, and that the ultrasound transducer itself has a capacitive output impedance ($C_{piezo}$). Thus, $R_b$ should be chosen so that 1/($R_b\cdot C_{piezo}$) is placed well above the bandwidth of $V_{neural}$ (>10 kHz) but well below the ultrasound frequency ($\sim$10 MHz for a 100 $\mu$m node). $R_b$ along with the transistor width must also be chosen carefully to achieve the best reflectivity, as will be described shortly.

Since modulation of $I_{DS}$ in turn modulates the impedance seen across the two piezoelectric drive terminals, the FET effectively modulates the backscattered signal seen by a distant transmitter. The change in the nominal level of $I_{DS}$ is a function of $V_{GS}$, which can be up to 10 $\mu$V ($V_{neural}$) for a 100 $\mu$m dust node near an active neuron. The sensitivity, S, to the action potential, then, is defined as the change in $I_{DS}$ with respect to $V_{GS}$ normalized by the nominal $I_{DS}$ (in addition to the current through $R_b$) and $V_{neural}$,
\begin{equation}
S=\mfrac{V_{neural}}{I_{DS}+V_{DS}/2R_b}\cdot\mfrac{\partial I_{DS}}{\partial V_{GS}}=V_{neural}\cdot\mfrac{g_{m}}{I_{DS}+V_{DS}/2R_b}
\end{equation}

Since $g_{m}$ (transconductance of a FET) is directly proportional to $I_{DS}$, in order to maximize $g_{m}/I_{DS}$ (i.e., achieves the largest $g_{m}$ for a given $I_{DS}$), we would like to operate the FET in its steepest region -- specifically, deep sub-threshold where it looks like a bipolar junction transistor (BJT). Therefore, the nominal $V_{GS}$ bias can be 0 V, which simplifies the bias circuitry. The modulation of the current is equivalent to a change in the effective impedance of the FET, or the electrical load to the piezoelectric transducer. This variation in the load impedance affects the ultrasonic wave reflectivity at the neural dust and modifies the wave that is backscattered. Note that in order to maximize the sensitivity (i.e., operating the transistor in deep sub-threshold), the system should be constrained such that the piezoelectric voltage is never too large compared to the threshold voltage.

A SPICE simulation of a typical low-threshold voltage NFET in a standard 65 nm CMOS technology was used in order to assess the nominal current level and the change in the effective impedance of the electrical load with $V_{neural}$. We assumed that we can implement suitably large $R_b$ in sufficiently small area of the neural dust nodes. As previously mentioned, in deep sub-threshold, the FET behaves as a BJT, where the physical limit on the achievable $g_{m}/I_{DS}$ = $q/k_BT$, determined by the Boltzmann distribution of carriers. As a result, we can obtain S = 400 ppm for $V_{neural}$ = 10 $\mu$V with a perfect BJT. Given the non-ideality factors associated with FETs, the sensitivity is reduced by a factor of 1.5 -- 2, to roughly 250 ppm, which is confirmed by the simulation.

The implication of the modification in the electrical properties of the NFET (output load of the piezoelectric transducer) on the change in the acoustic signal and the corresponding design specifications for the interrogator is discussed in detail below.

\section{System design and constraints: Interrogator}
\emph{Shorter transmission distance and larger aperture of the interrogator allow efficient trans-cranial power delivery via electromagnetics.}
\\
\\
The focus of the paper up to this point has been on the constraints associated with scaling the neural dust. In order to interface with the BMI electronics and to post-process recorded neural data for brain mapping, an interrogator that can extract the information of the sensor nodes, perform precise localization and addressing, and provide power for the communication needs to be designed. To achieve a BMI-relevant density of neural recordings, neural dust implants may need to be spaced as close as 100 $\mu$m (embedded up to a depth of 2 mm into the cortex). On the other hand, the interrogator elements will be larger than the sensor nodes and will be spaced at a larger pitch (between 100 $\mu$m and 1 mm). Furthermore, for the preliminary system, we assume that the interrogator is placed beneath the skull and below the dura mater, to avoid strong attenuation of ultrasound by bone ($\sim$22 dB/(cm$\cdot$MHz) \cite{Hoskins}) and to prevent wave reflection and efficiency loss from impedance mismatch between different tissue layers and the skull. The complete trans-cranial transmitter system then would nominally contain an EM link to couple information through the skull \cite{Sanni}. We do not discuss the design of the RF trans-cranial communication link as that is covered in other work.
\\
\\
\emph{Sufficient receiver sensitivity is required by the interrogator to resolve the occurrence of a neural spike.}
\\
\\
A different set of challenges exist in implementing circuitry to generate, collect and process neural data. Namely, innovative approaches are essential to 1) ensure that the interrogator/sensor combination has sufficient sensitivity to meet the necessary data resolution for BMI and 2) allow for combination of various multi-node interrogation strategies to distinguish among different sensor nodes.

For the analysis carried out in this paper, we assumed that the power and size constraints of the neural dust, and not the interrogator, are the major bottlenecks in the scaling of ultrasound-mediated neural dust system. In order to verify the validity of this assumption, we can examine, to the zeroth order, the power required by the interrogator to achieve certain receiver sensitivity for a passive implementation of the neural dust node. From the complete link model shown in {\bf Fig. 4}, we note that the change in the electrical impedance of the NFET load induces a change in the input admittance (or the input power) of the two-port network. The interrogator (receiver) must be able to detect this change in the input power level in order to resolve the occurrence of a neural spike. Therefore, we need to determine the size of the FET sensor on the dust node that maximizes the change in the input power level of the two-port network, or,
\begin{equation}
\Delta P_{in} \propto \left|\mfrac{Y_{in,spike}-Y_{in,nom}}{Y_{in,nom}}\right|
\end{equation}
where $Y_{in,spike}$ and $Y_{in,nom}$ denote input admittance of the two-port network with and without a neural spike, respectively. {\bf Fig. 8} shows the result of the optimization problem with a standard 65 nm CMOS technology. For 100 $\mu$m and 20 $\mu$m dust nodes, 75 $\mu$m and 16 $\mu$m width FET maximize $\Delta P_{in}$, respectively. Note that since the optimum transistor width (i.e., nominal impedance) for achieving the largest reflection is pretty flat, passive node is insensitive to the effects of threshold variability in the transistors and DC offsets in the neural electrodes.

The FET sensor design variable (transistor width), however, is constrained due to the thermal noise of the FET (which sets the lower limit) and the maximum available power at the node and the neural dust form factor (which set the upper limits). Clearly, the small footprint of the neural dust restricts the maximum effective width of the FET sensor that we can pack on the dust, and we term this the \emph{area limit}. More importantly, we need to ensure that the thermal voltage noise of the FET does not overwhelm the AP voltage. As a result, for a fixed bandwidth, in order to lower this voltage noise floor of the FET, it is necessary to increase the bias current, and hence the power consumption given a fixed output voltage. Given a simple single-ended transistor amplifier with a single dominant pole, a bias current of $I_{DS}$, and a transconductance of $g_m$, the minimum bias current required can be derived as,
\begin{equation}
I_{DS} = \mfrac{\pi}{4}\cdot\mfrac{4k_BT}{v_n^2}\cdot\mfrac{k_BT}{q}\cdot BW
\end{equation}
where $v_n^2$ is the input-referred voltage noise. As a result, the FET must be large enough to be able to sustain this minimum bias current. Therefore, for a BW = 10 kHz and voltage SNR at the input of the FET of 3 (which sets $v_n^2$ based on $V_{neural}$), we can compute the minimum allowable size of the FET, restricted by the \emph{noise limit}. Finally, in order to reliably operate the FET, the drain-source voltage of the FET must be at least $\sim$4 $k_BT/q$ or 100 mV. As a result, neural dust must capture enough power from the interrogator to sustain both 100 mV and the minimum current required to ensure that the thermal noise does not dominate the AP voltage. This is defined as the \emph{power limit}.
\begin{figure}[t]
\begin{center}
\centerline{\includegraphics[width=.49\textwidth]{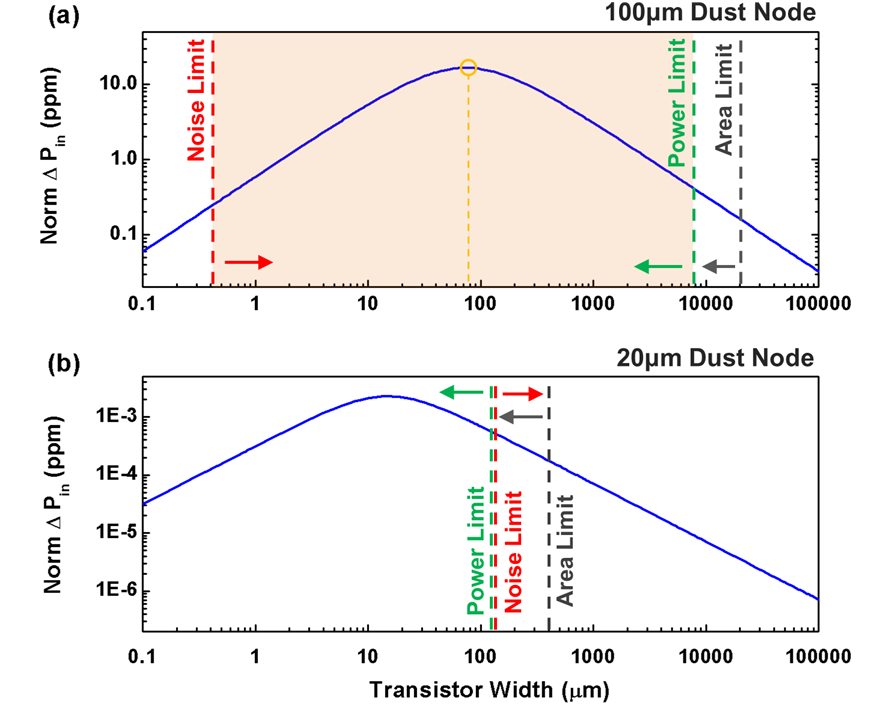}}
\caption{Change in the input power level (i.e., power at the interrogator) as a function of transistors width for a 65nm CMOS process and with {\bf (a)} 100 $\mu$m and {\bf (b)} 20 $\mu$m neural dust nodes.}
\label{fig6}
\end{center}
\end{figure}

With such restrictions, {\bf Fig. 8} shows that for a 100 $\mu$m dust node, we can design a FET sensor to generate a 16.6 ppm change in the input power with a measured $V_{neural}$. This results in $\sim$120 nW (-39 dBm) of backscattered power at the input given a 1 mm$^2$ interrogator aperture outputting 7.2 mW of power to satisfy safety regulations on output power density of 720 mW/cm$^2$. With such power levels, given a thermal noise spectral density of -174 dBm/Hz of input noise power, 10 kHz of BW, 10 dB of noise figure, and 10 dB of SNR, a traditional CMOS receiver should be sensitive enough to detect at minimum -114 dBm of input power. A number of highly-sensitivity receivers with < mW of DC power consumption have been demonstrated (e.g., \cite{Otis}).

For a 20 $\mu$m dust, however, {\bf Fig. 8} shows that the upper limit on the FET size imposed by the power limit is lower than the lower limit set by the noise limit, indicating that the passive implementation of neural dust system scales roughly to 20 $\mu$m.

\section{Re-design of neural dust node}
The scaling of both \emph{active} and \emph{passive} node implementations presented above is limited by the noise requirement of the front-end architectures, which is determined by the achievable differential signals between the electrodes. Decoupling the inherent tradeoff between the size of individual implants and the achievable SNR can improve the scaling of these implementations.
\\
\\
\emph{Re-thinking the design of neural dust can enhance its scalability.}
\\
\\
Since the trade-off derives directly not from the neural dust dimension, but from electrode separation, one approach may be to add very small footprint ($\sim$1 -- 5 $\mu$m wide) "tails" which position a single (or multiple) electrode relatively far (> 50 -- 100 $\mu$m) from the base of the neural dust implant. This would result in the design shown in {\bf Fig. 9}, where instead of placing a single differential surface electrode on neural dust, the neural dust can consist of a short strand of flexible and ultra-compliant substrate populated with recording sites. Assuming that the achievable electrode separation in the tail of a 20 $\mu$m node is 100 $\mu$m, this implies that the noise limit, as shown in {\bf Fig. 8}, will set the lower bound to 0.4 $\mu$m of transistor width and allow the design of a FET sensor on the dust node that achieves the optimal sensitivity, at 2.3e-3 ppm. This corresponds to 16.6 pW (-77.8 dBm) of backscattered power at the input, which is still in the realm of feasibility with a traditional CMOS receiver \cite{Otis}. Therefore, this approach can address one of the major pitfalls with only a minor adjustment to the original idea as this neural dust still operates under the same principle as before, but has higher achievable SNR.

Note that the exact technology used for the previous analysis is not critical to the conclusion we drew. Although the absolute value of the impedance level is important since it determines the reflection coefficient, and therefore, the efficacy of the backscatter, as shown in {\bf Fig. 8}, the analysis above indicates that the optimal transistor width for the maximal sensitivity is small compared to the available neural dust footprint. Therefore, although the threshold voltage (hence the nominal impedance level per transistor width) may vary among different technology nodes, achieving the optimal impedance level within the footprint may not be an issue.
\begin{figure}[t]
\begin{center}
\centerline{\includegraphics[width=.54\textwidth]{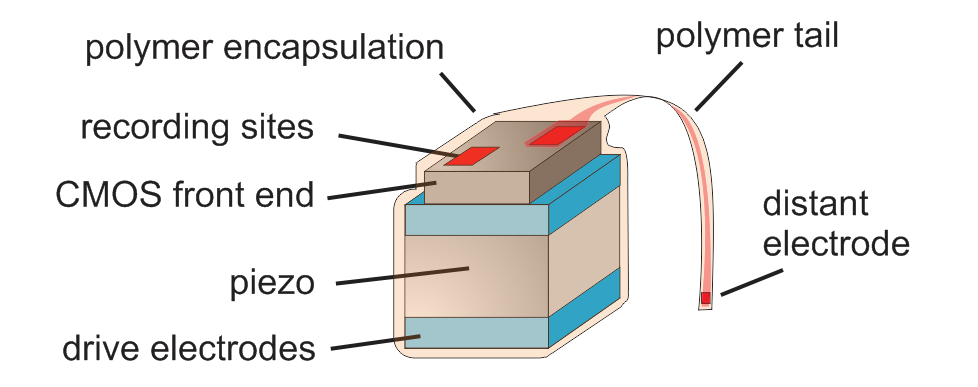}}
\caption{Neural dust with an ultra-compliant flexible polyimide "tail", populated with recording sites, can be envisioned to bypass the limits of the achievable differential signal between two electrodes placed on a neural dust footprint.}
\label{fig6}
\end{center}
\end{figure}

In addition, since the analysis above does not take into accounts additional interference (e.g., ultrasonic wave reflection from other structures in the brain, such as vasculature), the sensitivity requirement of the interrogator are more stringent than predicted earlier. Such reflections will likely lead to intersymbol interference. In the case of an active node, such interference can be dealt with through adaptive equalization and/or error correcting codes \cite{Proakis}. For the passive system -- which is effectively "transmitting" analog information back to the interrogator through the backscatter -- some form of filtering could be applied to reverse the effects of these reflections. Alternatively, one could potentially utilize a pulse-based system to uniquely discriminate the various reflections based on their arrival times. 

\section{Discussion and Conclusions}
The analysis presented points to three major challenges in the realization of ultra-small, ultrasound-based neural recording systems. The first is the design and demonstration of front-ends suitable for operating within the extreme constraints of decreasing available power and decreasing SNR with scale. This could be addressed with a combination of CMOS process and design innovation as well as thinned, multi-substrate integration strategies (see, for example, \cite{Sillon, Smith}). The second challenge is the integration of extremely small piezoelectric transducers and CMOS electronics in a properly encapsulated package. The above discussion assumed the entire neural dust implant was encapsulated in an inert polymer or insulator film (a variety of such coatings are used routinely in neural recording devices; these include parylene, polyimide, silicon nitride and silicon dioxide, among others) while exposing two recording electrodes to the brain. The addition of "tails" as discussed above presents additional fabrication challenges. The third challenge arises in the design and implementation of suitably sensitive sub-cranial transceivers which can operate at low power (to avoid heating between skull and brain). In addition to these three challenges, this paper does not discuss \emph{how} to deliver neural dust nodes into the cortex. The most direct approach would be to implant them at the tips of fine-wire arrays similar to those already used for neural recording. Neural dust nodes would be fabricated or post-fab assembled on the tips of array shanks, held there by surface tension or resorbable layers; a recent result demonstrates a similar approach to implant untethered LEDs into neural tissue \cite{Kim}. Once inserted and free, the array shanks would be withdrawn, allowing the tissue to heal. Kinetic delivery might also be an option, but there is no existing data to evaluate what effect such a method would have on brain tissue or the devices themselves.

The trans-cranial transmitter design also introduces multi-interrogator, multi-node communication possibilities that will need to be developed in order to enable the large number of recording sites envisioned in this paper. Because the neural dust nodes are smaller than a wavelength, the reflected signals will be subject to diffraction. With multiple nodes embedded and sufficiently wide transceivers, this presents an interesting inverse problem of potential benefit in resolving signals from different nodes. An alternative approach to multi-node communication would be to fabricate nodes with a variety of resonant frequencies and use frequency discrimination (i.e., each dust transmits on its own frequency channel). Lastly, neural dust nodes with aspect ratios close to 1:1:1 will not only couple energy into modes along the two axes perpendicular to the transmission axis, they will also re-radiate along those axes. This means nodes lying near each other on a "horizontal" plane (relative to the top surface of the cortex) may see inter-node signal mixing. This has interesting implications for node-to-node communication.

Lastly, one of the more compelling possibilities would be to harness the considerable volume of research that has gone into micro- and nanoelectromechanical RF resonators (which easily operate in the MHz range \cite{Sadek, Lin} and thin-film piezoelectric transducers \cite{Przybyla, TM} to produce devices with better power coupling as a function of scale, thus facilitating extremely small (10's of $\mu$m) dust nodes. This remains an open opportunity.

\begin{acknowledgments}
The authors would like to thank Tim J. Blanche of Allen Institute of Brain Science, Konrad P. Kording of Northwestern University, Adam H. Marblestone of Harvard University, Emmanuel Quevy of Silicon Laboratories, Mikhail G. Shapiro of California Institute of Technology, Bradley M. Zamft of the US Department of Energy, and William Biederman, Peter Ledochowitsch, Nathan Narevsky, Christopher Sutardja, and Daniel J. Yeager of UC Berkeley for valuable discussions. This work was supported by the NSF Graduate Fellowship for DS and the Bakar Fellowship for JMC and MMM.
\end{acknowledgments}

\end{article}
\end{document}